\documentclass[10pt]{article}

\usepackage{makeidx}
\usepackage{epsfig}
\usepackage{float}
\usepackage{amscd}
\usepackage{amsmath}
\usepackage{amssymb}
\usepackage{color}
\usepackage{fancyheadings}
\usepackage{fancyvrb}
\usepackage{slashbox}
\usepackage{caption}
\usepackage{subcaption}

\newcommand{\pfn}{{\phi}}
\newcommand{\sfn}{{\psi}}

\newcommand{\btau}{{\boldsymbol \tau}}
\newcommand{\omitit}[1]{}
\newcommand{\Reyuls}{\mathbb{R}}

\newcommand{\p}{{{\rm I}{\relax\ifmmode\mskip-\thinmuskip\relax\else\kern-.22em\fi}{\bf P}}}
\newcommand{\perpzh}{\perp\kern-.20em^a}

\newcommand{\Plang}{{\p}{\kern-.03em{\rm fortran}}}
\newcommand{\Pfortran}{{\p}{\kern-.03em{\rm fortran}}}

\newcommand{\half}{{\textstyle{1\over 2}}}

\newcommand{\set}[2]{\left\lbrace #1 \; : \; #2 \right\rbrace}

\newfloat{code}{ht}{aux}[section]
\floatstyle{boxed}
\floatname{code}{Program}
\restylefloat{code}

\newtheorem{theorem}{Theorem}[section]
\newtheorem{lemma}{Lemma}[section]

\newcommand{\norm}[1]{\Vert{#1}\Vert}

\newcommand{\tbnorm}[1]{\vert\kern-.1em\vert\kern-.1em\vert{\, #1 \,}\vert\kern-.1em\vert\kern-.1em\vert}

\newcommand{\oput}[1]{\rlap{${}{\lower 1.3ex\hbox{${\sim}$}}$}{#1}}

\newcommand{\sdiv}{{{\nabla\cdot} \,}}
\newcommand{\curl}{{{\rm curl} \,}}

\newcommand{\bfz}{{\mathbf 0}}

\newcommand{\uu}{{\mathbf u}}

\newcommand{\vv}{{\mathbf v}}

\newcommand{\gbc}{{\mathbf g}}

\newcommand{\nn}{{\mathbf n}}
\newcommand{\ff}{{\mathbf f}}
\newcommand{\xx}{{\mathbf x}}

\newcommand{\datekern}{{\relax\kern+.1em}--{\relax\kern-.01em}}

\newcommand{\antiparallel}{\not\kern-0.35em\Vert}

\newcommand{\order}[1]{{\cal O}\left(#1\right)}

 \title{New Insights on the Stokes Paradox for Flow in Unbounded Domains}
 \author{Ingeborg G. Gjerde and L. Ridgway Scott}

\begin{document}

\maketitle

\begin{abstract}
Stokes flow equations, used to model creeping flow, are
 a commonly used simplification of the Navier--Stokes equations. The simplification is valid for flows where the inertial forces are negligible compared to the viscous forces. In infinite domains, this simplification leads to a fundamental paradox.

In this work we review the Stokes paradox and present new insights related to recent research. We approach the paradox from three different points of view: modern functional analysis, numerical simulations, and classical analytic techniques. The first approach yields a novel, rigorous derivation of the paradox. We also show that relaxing the Stokes no-slip condition (by introducing a Navier's friction condition) in one case resolves the Stokes paradox but gives rise to d'Alembert's paradox.

The Stokes paradox has previously been resolved by Oseen, who showed that it is caused by a limited validity of Stokes' approximation. We show that the paradox still holds for the Reynolds--Orr equations describing kinetic energy flow instability, meaning that flow instability steadily increases with domain size. We refer to this as an instability paradox.
\end{abstract}

\section{Introduction}

Fluids like air and water can be modeled to high precision using the Navier--Stokes equations \cite{temamns}.
These equations are still intensively studied, and they have some very curious mathematical features, often described as paradoxes, such as d'Alembert's
paradox \cite{lrsBIBjn,stewartson1981d} and 
Whitehead's paradox \cite[section 8.3]{vandykeperturbmeth}.
Here we discuss one of the most well known, namely the Stokes paradox.
The Stokes paradox also arises for more exotic fluids, such as Fermi electrons \cite{PhysRevB.95.115425}.

Consider a domain $\Omega \subset \mathbb{R}^3$ containing an infinitely long cylinder with radius 1. The cylinder has a boundary denoted $\Gamma$. The domain is filled with a fluid moving with velocity $\uu$ and pressure $p$ past the cylinder. Mathematically, this can be described using the Navier--Stokes equations
\begin{equation} \label{eqn:effnavstot}
\begin{split}
-\Delta \uu + \nabla p &=-R \,\uu\cdot\nabla \uu \quad \hbox{in}\;\Omega,\\
\sdiv\uu &=0 \quad \;\hbox{in}\;\Omega, \\
\end{split}
\end{equation}
together with the boundary condition
\begin{equation} \label{eqn:bceesnavst}
\uu=\gbc\;\hbox{on}\;\Gamma.
\end{equation}
\omitit{
The first and second equations describe conservation of momentum and mass, respectively. The conservation of momentum arises from applying Newton's second law ($F=ma$) to balance the forces acting on the fluid. The forces at work are due to fluid viscosity (captured by the $-\Delta \uu$-term), fluid pressure (captured by the $\nabla p$-term) and fluid momentum (captured by $R \,\uu\cdot\nabla \uu$). The latter term is non-linear, which makes the system more difficult to approximate (requiring e.g. a Newton method) and rather non-trivial to analyze.
}

In this form, the Navier--Stokes equations are posed with only one parameter, 
namely the Reynolds number $R$ \cite{LandauLifshitz}:
\begin{equation} \label{eqn:renodefvst}
R=U L / \nu.
\end{equation}
Here, $U$ is the representative flow velocity, $L$ the representative length for
the domain, and $\nu$ is the kinematic fluid viscosity. 

\omitit{
The Reynolds number can be thought of as the ratio of inertial forces to viscous forces within the fluid. For a fluid with low viscosity, the Reynolds number will be high, and the inertial forces dominate. If we think of tracking two fluid particles with different velocity and direction, their momentum will keep them moving and they may crash into each other and change direction. As the Reynolds number increases, the flow can therefore develop different forms of turbulence. Thinking conversely of the Reynolds number decreasing, the fluid becomes more syrupy, meaning that viscous forces dominate. In this case the flow becomes laminar.  

\begin{remark}[Calculating the Reynolds number]
\label{rem:reynolds-number}
It is not always clear how to choose the flow velocity $U$ and length scale $L$ when calculating the Reynolds number.
Close friends can differ on the choices. For the flow velocity $U$ one typically picks the maximal or average flow rate. The choice affects the Reynolds number by a numerical factor.
More complicated is the representative length scale. For flow around an obstacle, it is common to pick a dimension describing the object. But for a cylinder, do you pick the radius or the diameter? Here the convention is to use the diameter. In any case, it is important to be clear about how the Reynolds number was chosen, so that different simulations (and experiments) can be compared reliably.
\end{remark}
}

In this work, we are interested in the case where the Reynolds number will be small, and we have $R \ll 1$. In this case, it seems reasonable to drop the advective term, in which case we have the simpler Stokes equations
\begin{equation} \label{eqn:firstnavst}
\begin{split}
-\Delta \uu + \nabla p &= 0  \;\hbox{in}\;\Omega,\\
\sdiv\uu &=0\;\hbox{in}\;\Omega,
\end{split}
\end{equation}
again with a boundary condition of the form \eqref{eqn:bceesnavst}. 

If the domain $\Omega$ is taken to be infinitely large, however, this simplification leads to a fundamental paradox \cite{stokes1851effect, ref:FinnExterioreview}. 
The Stokes paradox is that ``creeping flow of an incompressible Newtonian fluid around 
a cylinder in an unbounded fluid has no solution'' \cite{ref:tannerpowerlawparadox}.
Others have characterized the paradox by saying
\begin{itemize}
\item
\cite[1st paragraph]{shaw2009simple}
``Stokes (1851) established that there was no solution to the two-dimensional, steady,
incompressible, Navier--Stokes equations for asymptotically uniform flow around a cylinder.''
\item
\cite[Remark 3.15]{ref:wieghtedSobolevStokes} ``no classical solution ...
that tends to a nonzero vector at infinity.''
\end{itemize}
These statements require some clarification.
Firstly, potential flow around a cylinder solves the Stokes equations
with the right boundary conditions at infinity.
However, potential flow (Figure \ref{fig:potflow}) does not satisfy no-slip
boundary conditions on the cylinder;
for this reason it has commonly been discarded as physically incorrect.

\omitit{
\begin{figure}
\centering
\includegraphics[width=0.25\textwidth]{cylinder}

\vspace{1em}
\begin{subfigure}{0.49\textwidth}
\includegraphics[width=1\textwidth]{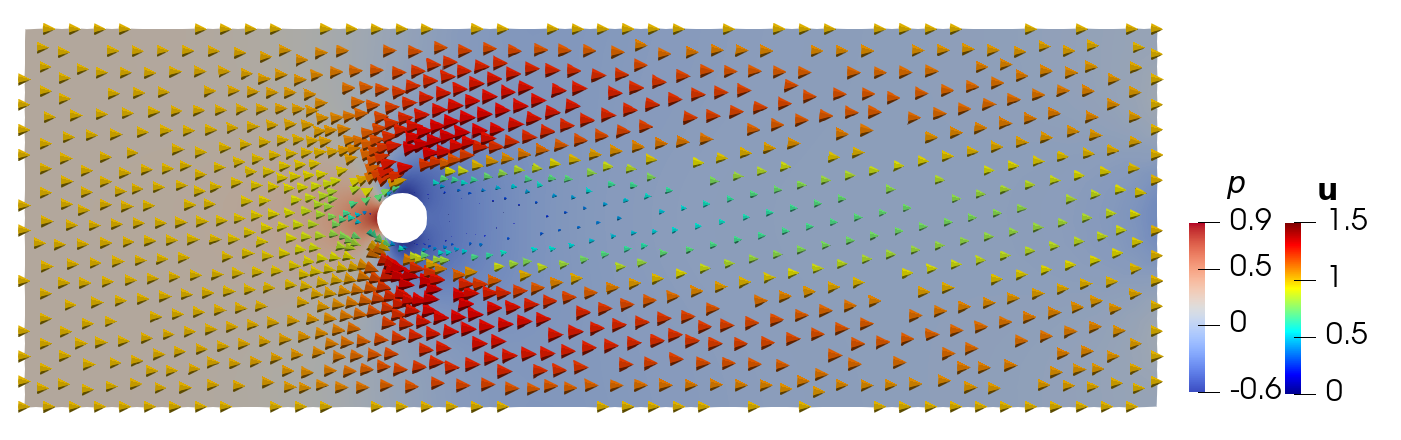}
\caption{Navier--Stokes flow with $\nu=0.1$}
\label{fig:ns-nu0}
\end{subfigure}
\begin{subfigure}{0.49\textwidth}
\includegraphics[width=1\textwidth]{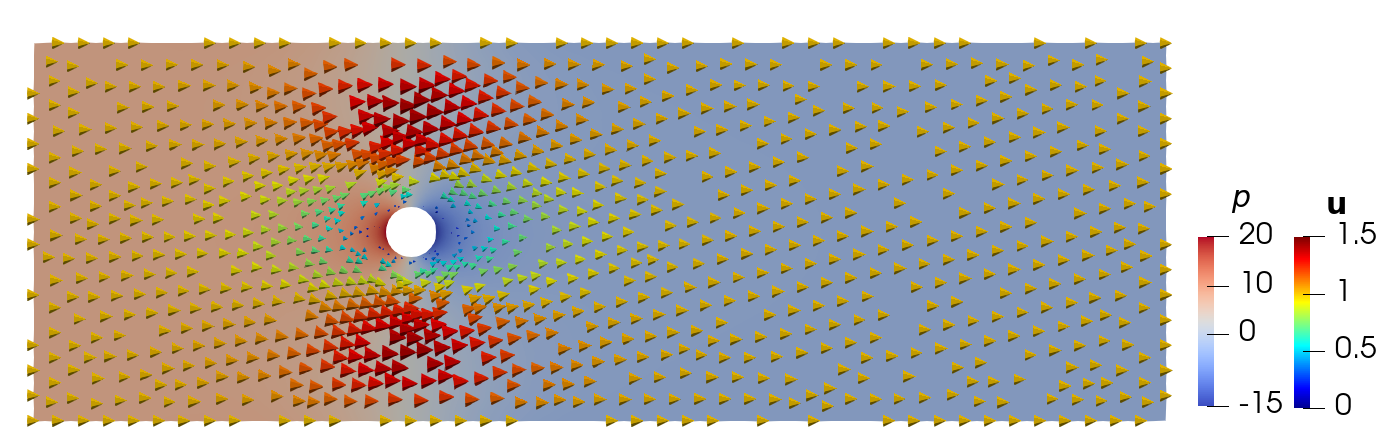}
\caption{Navier--Stokes flow with $\nu=10$}
\label{fig:ns-nu1}
\end{subfigure}
\begin{subfigure}{0.49\textwidth}
\includegraphics[width=1\textwidth]{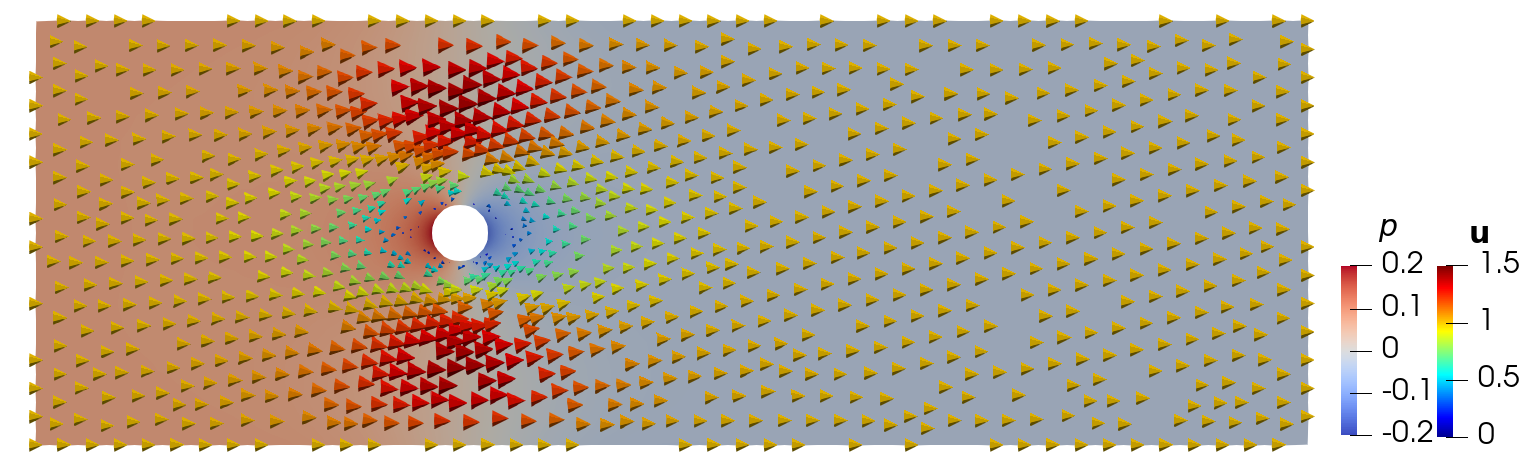}
\caption{Stokes flow with $\nu=0.1$}
\label{fig:s-nu0}
\end{subfigure}
\begin{subfigure}{0.49
\textwidth}
\includegraphics[width=1\textwidth]{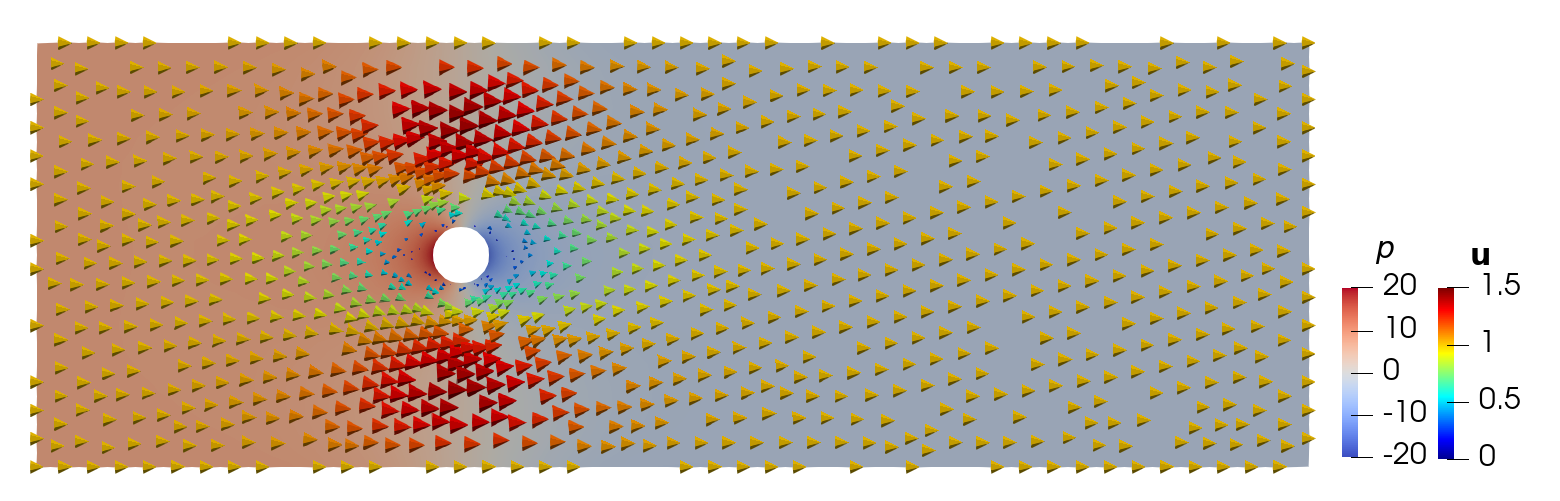}
\caption{Stokes flow with $\nu=10$}
\label{fig:s-nu1}
\end{subfigure}
\caption{Stokes and Navier-Stokes equations solved in a rectangular domain $\Omega$ containing a circle of radius 1 with boundary $\Gamma$. The flow is assigned a uniform flow profile $\uu=(1,0)$ on the sides of the box and a no-slip boundary condition on $\Gamma$ ($\uu=(0,0))$). For $\nu=0.1$, the fluid has a comparatively high Reynolds number. We therefore see distinct differences in the Navier-Stokes flow in Figure \ref{fig:ns-nu0} and Stokes flow in Figure \ref{fig:s-nu0}. For the higher viscosity $\nu=10$, the fluid has a comparatively low Reynolds number. In this case we only see small differences in the Navier-Stokes flow in Figure \ref{fig:ns-nu1} and Stokes flow in Figure \ref{fig:s-nu1}.}
\end{figure}
}

To handle the no-slip boundary condition we instead look to the literature on functional analysis. A precise mathematical theory for the Stokes equations in an unbounded domain was developed in \cite{girault1991well}. 
We can explain the results there informally as follows: 
That it is a well posed problem to require a reasonable flow profile around a reasonable domain. A solution therefore \textit{exists} solving Stokes equations in an infinite domain enclosing a cylinder, as long as we set reasonable boundary conditions on the cylinder.
But the paradox is that we cannot set a boundary condition at infinity.

This point is often misunderstood, saying instead that there is a 
``need to satisfy two boundary 
conditions, one on the object and one at infinity'' \cite{badref}.
This is despite the fact that many papers have dealt with the paradox and its resolution.
The first resolution was by Oseen \cite{oseen1927neuere} who realized that it was
necessary to keep $R>0$ in \eqref{eqn:effnavstot}
and provided an approximate (linearized) solution to Navier-Stokes.
This observation has been substantially amplified by Finn \cite{ref:FinnExterioreview}.
But many others have dealt with the paradox, for example by improving the Oseen
approximation \cite{ref:KaplunLagerstromSparadox,ref:ProudmanPearssonSparadox}.
For a survey and history of their methods, see \cite[Chapter V]{vandykeperturbmeth}
where the so-called matched asymptotic expansions are traced back to seminal work
by K.~O.~Friedrichs.

The Sobolev space in which the solution
exists provides important context for the Stokes paradox. To be more precise, the solution exists in a special weighted Sobolev space in which we give up control over the function as we get infinitely far from the cylinder. For this reason it is not possible to prescribe the flow at infinity. We will show that, if the flow is specified to be a constant on the cylinder, the flow must be that constant everywhere. Thus, while a non-trivial solution \textit{exists}, it may not be physically meaningful.

Due to the confusion related to the Stokes paradox, we examine it in some detail
before proceeding to the main results of the paper.
We draw together disparate approaches to give the fullest possible understanding.
Although much of what we present at the beginning can be found separately elsewhere, 
the combination of the different approaches gives a more complete
picture than has so far been presented in one place.

The article will proceed as follows. We begin in Section \ref{sec:classical-sols}
by giving two analytic solutions for Stokes flow around a cylinder and discuss their properties. 
Next, we give in Section \ref{sec:stokesparadox} a derivation of the Stokes paradox,
using the mathematical framework from \cite{girault1991well}.
In Section \ref{sec:navrevenge}, we examine the impact of a different boundary 
condition on the cylinder, Navier's slip (friction) condition.
We see that it is possible to resolve the Stokes paradox with one value of the
friction parameter, although that value seems to be nonphysical.

We next view the Stokes paradox through other lenses.
In particular, we solve the Stokes problem on a very large (but finite) domain 
surrounding the cylinder.
In this case we can pose any boundary conditions we like on the outer boundary. So what goes wrong as we let the outer boundary go to infinity? 
In Section \ref{sec:bounded-domains} we answer this in two ways, using modern computational methods (Section \ref{sec:compapp}) and classical analytical solution techniques (Section \ref{sec:analytic}). We will see that they give the same answer: the solution goes to a constant. Thus we have multiple ways of viewing the Stokes paradox, each with its own advantages and limitations.

In Section \ref{sec:nsnopar} we review the resolution of the Stokes paradox 
by Oseen \cite{oseen1927neuere}.
According to Oseen, the paradox is caused by the limited validity of 
Stokes' approximation, which relies on the Reynolds number being small.
To be more precise, we have two limits going to infinity: the viscosity
and the domain size. 
Either limit alone is well behaved, but jointly they are not.
Thus one must consider the full Navier-Stokes equations if the domain is large. We also discuss briefly some open questions regarding the existence of a solution for the Navier--Stokes equations when the Reynolds number grows large.

In Section \ref{sec:extox} we describe extensions of the Stokes paradox
concepts to other flow problems. In particular, we discuss how the Stokes paradox relates to flow instabilities. The Reynolds-Orr method gives a way to calculate the kinetic energy instability of a perturbed base flow, where the most unstable flow perturbations are calculated by solving a Stokes type eigenvalue problem. In \cite{lrsBIBiw}, it was observed that the instability of a base flow kept increasing as the domain grew. This can be explained as a particular instance of the Stokes paradox. To the best of our knowledge, however, it cannot be resolved in any such way as Oseen's resolution of the Stokes paradox.

\section{Classical solutions}
\label{sec:classical-sols}

\begin{figure}
\begin{subfigure}{0.41\textwidth}
\includegraphics[width=1.25\textwidth]{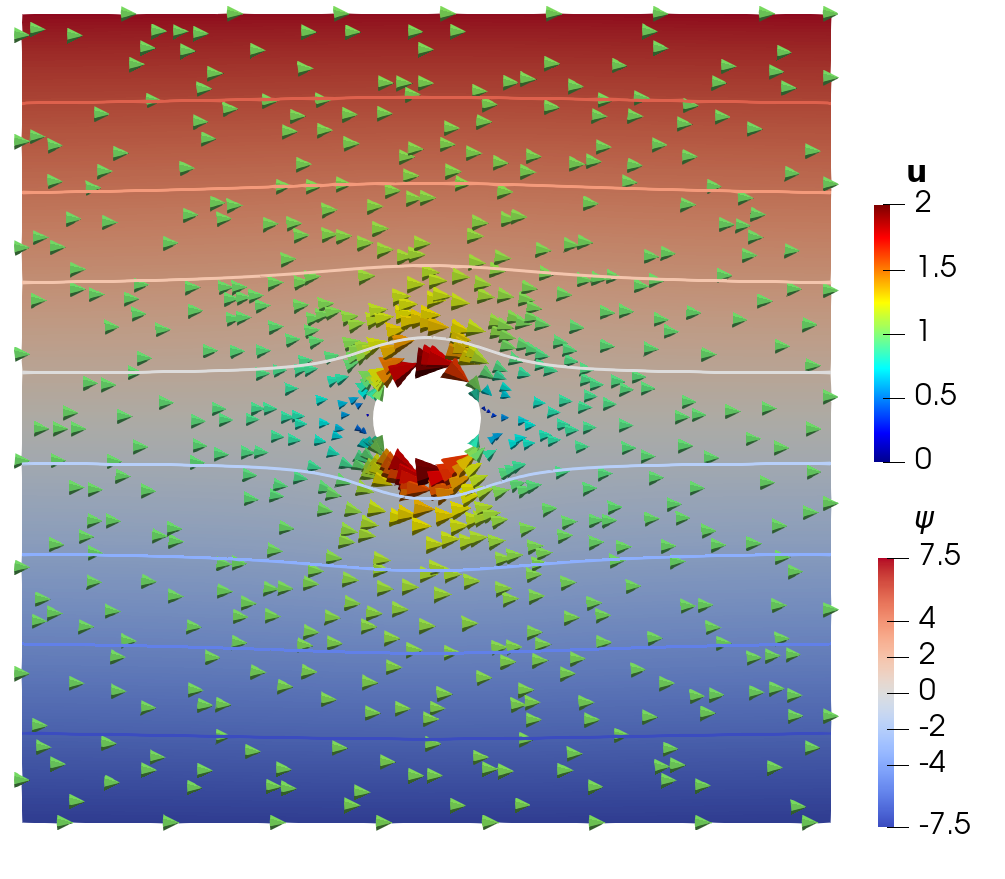}
\caption{Potential flow, $\beta=-2\nu$}
\label{fig:potflow}
\end{subfigure}\hspace{11mm}
\begin{subfigure}{0.45\textwidth}
\includegraphics[width=1.13\textwidth]{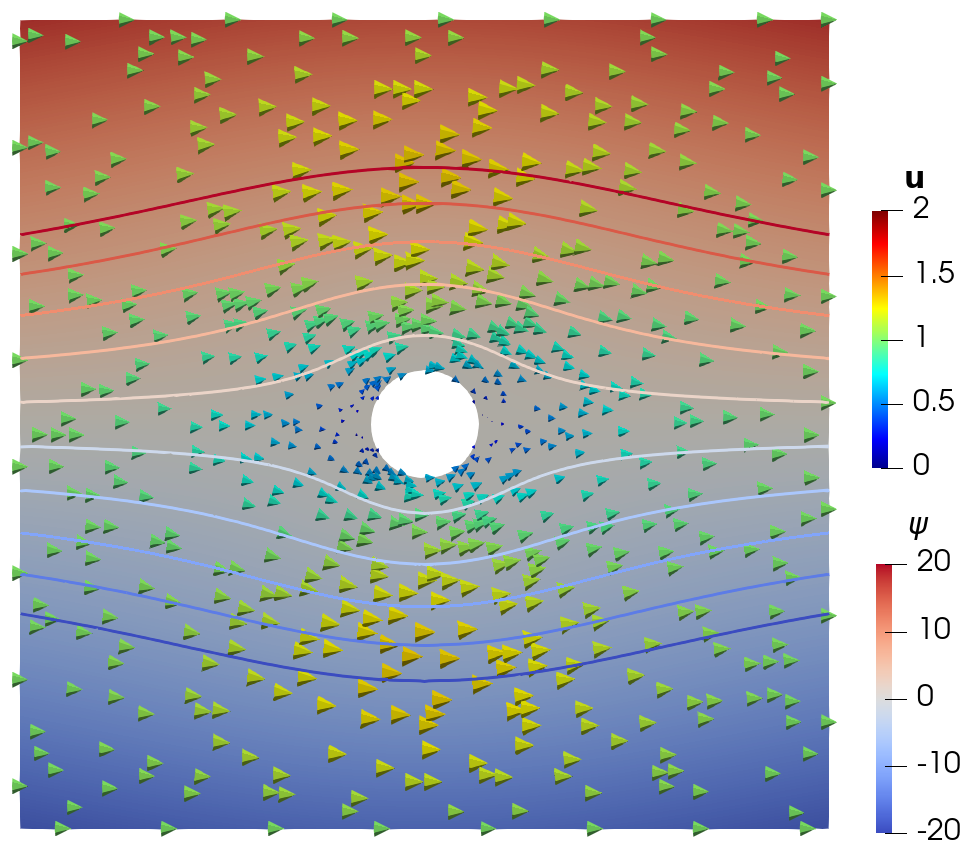}
\caption{Zero friction flow, $\beta=0$}
\label{fig:freeflow}
\end{subfigure}
\caption{Stream function $\sfn$, stream lines and flux $\uu$ (cones) for two classically known analytic solutions of the Stokes equations for flow past an infinitely long cylinder. Neither solution satisfies the no-slip boundary condition.}
\end{figure}

Assume for the moment $\Gamma$ to be an infinitely long cylinder of radius 1 in the $z$-direction with origin $(0,0)$ in the $x,y$-plane. Consider now the Stokes flow equations \eqref{eqn:firstnavst} with boundary condition \eqref{eqn:bceesnavst}. We now construct two analytic solutions. The strategy is to find a function $\uu$ that is divergence free so it satisfies the second equation in \eqref{eqn:firstnavst}. If the Laplacian of $\uu$ is a gradient of some function, we can solve the first equation in \eqref{eqn:firstnavst} by construction by taking the pressure equal to said function.

To find a function $\uu$ that is divergence free, we use two different approaches.
For the first, we take $\pfn$ to be a potential function, 
i.e.~we set $\uu=\nabla \pfn=(\pfn_x,\pfn_y)$, where $\pfn_i$ denotes the derivative of $\pfn$ in the $i$th direction. Then if we want $\nabla\cdot \uu=0$ we want $\nabla \cdot \nabla \pfn=\Delta \pfn=0$, i.e.~$\pfn$ to be harmonic.
Many harmonic functions are known in the literature; we take
\begin{equation} \label{eqn:sokesolst}
\pfn(r,\theta)=-(\cos\theta)\Big(r-\frac1r\Big)=-x\Big(1-\frac{1}{r^2}\Big).
\end{equation}
A straightforward calculation shows that $\pfn $ is harmonic.

We also have
\begin{align*}
\Delta\uu&=(\Delta \pfn_x, \Delta \pfn_y)=((\Delta \pfn)_x, (\Delta \pfn)_y) = 0.
\end{align*}
Thus we have a solution to the Stokes equation \eqref{eqn:firstnavst} if the pressure is taken to be a constant (so that $\nabla p=0$ as well).

Differentiating $r=\sqrt{x^2+y^2}$ we find
\begin{equation}\label{eqn:chacoronly}
r_x=\frac{x}{r},\qquad r_y=\frac{y}{r}.
\end{equation}
Using these, we have
$$
\pfn_x=1-r^{-2}+2yr^{-3}r_y
=1+(y^2-x^2)r^{-4}, \quad
\pfn_y=2yr^{-3}r_x=2xyr^{-4}.
$$
Then
\begin{equation} \label{eqn:yousokesolst}
\uu(x,y)=(\pfn_x(x,y),\pfn_y(x,y))=\Big(
1+\frac{y^2-x^2}{(x^2+y^2)^2},\frac{-2xy}{(x^2+y^2)^2}\Big).
\end{equation}

This solution for $\uu$ is commonly referred to as \textit{potential flow}. The potential flow solution is shown in Figure \ref{fig:potflow} together with its stream function and streamlines. We see that $\uu$ goes to uniform flow at infinity. Moreover, $\uu\cdot\nn=0$ on the cylinder. However, the tangential velocity $\uu\cdot\btau\neq 0$, meaning that this solution does not satisfy a no-slip boundary condition on $\Gamma$. This led Stokes to reject this solution.

Let us therefore make another analytic solution, this time trying the second approach to making a divergence-free flux $\uu$. 
By augmenting our vectors with a $z$-component we can define $\uu$ to be the curl of a vector $(0,0,\sfn)$ so that $\uu=(\sfn_y, -\sfn_x, 0)$. Dropping the $z$-coordinate again we have $\uu=(\sfn_y, -\sfn_x)$ and $\sdiv\uu=\sfn_{yx}-\sfn_{xy}=0$ as long as $\sfn$
is sufficiently smooth\footnote{for example $\sfn \in C^2(\Omega)$, i.e. two times continuously differentiable. In this case we can change the order of differentiation without issues.}.
For example, define
\begin{align}
\sfn=(\sin\theta)\, r\log r =y\log r \label{eq:stream-2}.
\end{align}
Then
\begin{align} \label{eqn:zerofriction}
\uu(x,y) = \Big(\frac{y^2}{r^2} + \log(r), -\frac{xy}{r^2}\Big).
\end{align}

By construction $\uu$ satisfies the second equation in \eqref{eqn:firstnavst}. The first equation in \eqref{eqn:firstnavst} can then be satisfied by choosing $p$ so that $\nabla p = \Delta \uu$. We will show in Section \ref{sec:analytic} that such a solution $p$ exists.

Again, the solution does not satisfy the no-slip boundary condition on $\Gamma$. Moreover, the solution diverges to infinity as $r\rightarrow \infty$. 

In the next section, we see how these analytic solutions are related to the Stokes paradox.

\section{Derivation of Stokes paradox using the variational framework}
\label{sec:stokesparadox}

\omitit{\begin{figure}
\centering
\includegraphics[width=0.45\textwidth]{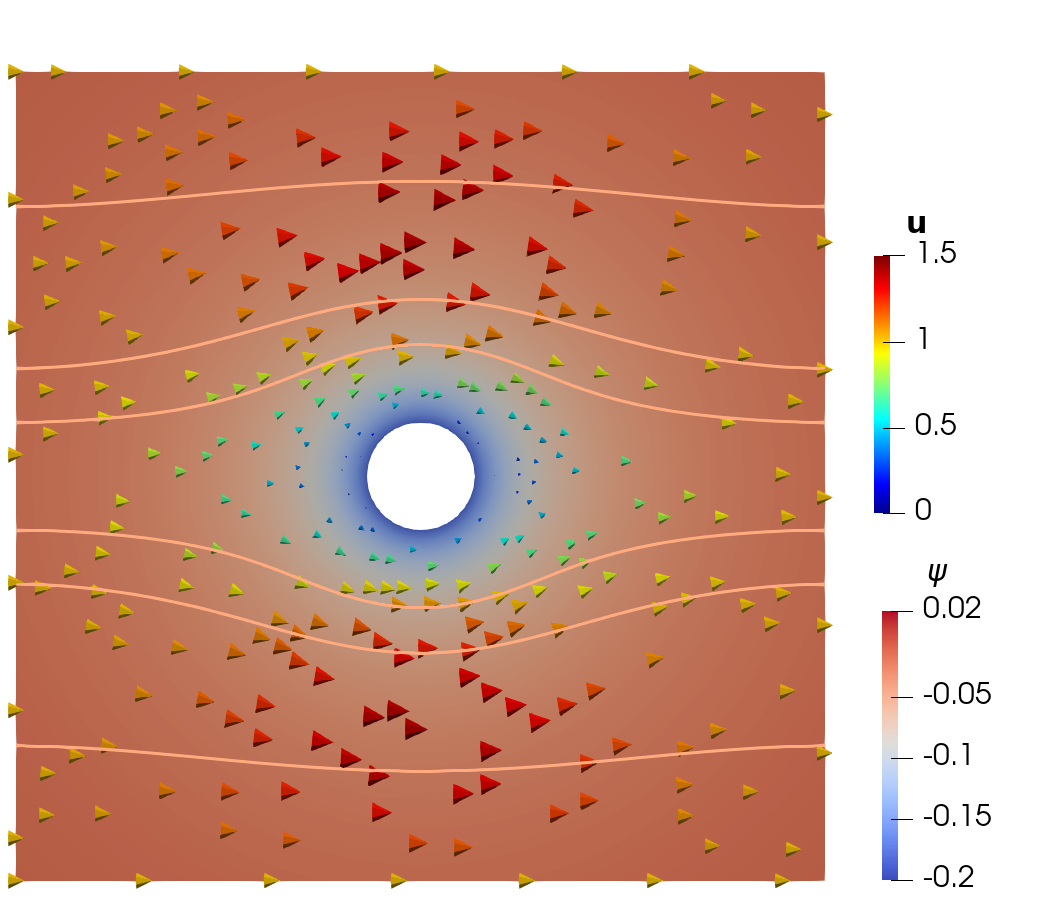}
\label{fig:noslip}
\caption{Stream function $\psi$, stream lines (orange lines) and flux $\uu$ (cones) for the solution of the Stokes equation with no-slip boundary conditions. The stream function for this flow will be derived in Section \ref{sec:alg-pdes}}.
\end{figure}}
 
The Stokes paradox occurs for incompressible flow Stokes flow with no-slip boundary conditions on the cylinder (i.e. $\mathbf{g}=0$ in \eqref{eqn:bceesnavst}). The paradox can be derived in several ways. In this work, we present three approaches: (i) a rigorous derivation using weighted Sobolev spaces, (ii) a formal approach using simulations in domains of increasing size and (iii) a semi-formal approach of deriving analytic solutions in bounded domains and passing to the limit.

\subsection{Sobolev spaces for the Stokes equations}
\label{sec:sobolev}

\omitit{
Sobolev spaces have become the workhorses for the modern approach to partial differential equations (PDEs). A Sobolev space is a type of Hilbert space where one measures the size of a solution and its derivatives. They can appear daunting at first; in this work we will focus on the main ideas, leaving further details to more advanced texts \cite{brennerscott}.

The well posedness of \eqref{eqn:firstnavst} depends on the domain $\Omega$.
}

If the domain $\Omega$ is 
Lipschitz continuous \cite[section 5.1]{giraultraviart}), the system is well posed 
with $\uu \in (H^1(\Omega))^d$, $d\in \{2,3\}$ being the dimension of the domain,
and $p \in L^2(\Omega/\Reyuls)$, where
\begin{align*}
L^2(\Omega)/\Reyuls= \set{ v : \Omega \rightarrow \mathbb{R}}{
 \int_{\Omega} v^2 \,d\xx < \infty, \quad \int_{\Omega} v \,d\xx =0 }
\end{align*}
is the space of square-integrable functions,
together with the norm given by 
$$
\norm{v}_{L^2(\Omega)}=\sqrt{\int_\Omega v^2 \, d\xx}.
$$
We also define
\begin{align*}
H^1(\Omega)= \set{ v \in L^2(\Omega)}{ \nabla v \in L^2(\Omega) }
\end{align*}
and the norm
\begin{equation*}
\norm{v}_{H^1(\Omega)}=\sqrt{\int_\Omega|\nabla v(\xx)|^2
          +v(\xx)^2\,d\xx},
\end{equation*}
where $\vert v(\xx)\vert$ is the Euclidean norm of $\nabla v(\xx)$. The notation $\uu \in (H^1(\Omega))^d$ then means that every component of $\uu$ is in $H^1(\Omega)$. To simplify notation, we from now on drop the superscript and simply write $\uu \in H^1(\Omega)$.

In summary, given a bounded domain and reasonable $\ff$ and $\mathbf{g}$ (for example $\ff \in L^2(\Omega)$ and $\mathbf{g} \in H^1(\Omega)$ \cite{giraultraviart}) there exists a unique solution pair $\uu \in H^1(\Omega)$ and $p\in L^2(\Omega)$ of \eqref{eqn:firstnavst}. Once we know there exists a unique solution, we can use numerical methods to solve for its approximation.

If the domain $\Omega$ is infinite, the previous result no longer holds.
Instead, the Stokes equation will be well posed with $p\in L^2(\Omega)/\Reyuls$
and $ \uu \in H^1_w(\Omega)$ defined by the norm
\begin{equation} \label{eqn:dubyahonet}
\norm{\vv}_{H^1_w(\Omega)}=\sqrt{\int_\Omega|\nabla\vv(\xx)|^2
          +\big(1+|\xx|\log |\xx|\big)^{-2}|\vv(\xx)|^2\,d\xx}.
\end{equation}
Centrally, this is a \textit{weaker} norm than the one we had for the $H^1(\Omega)$. Both spaces require the gradient of the function itself to be square-integrable, but in the $H^1_w(\Omega)$-space we only require the function to be square-integrable when multiplied by a weight function $\big(1+|\xx|\log |\xx|\big)^{-1}$. As this weight function goes to zero as $\xx \rightarrow \infty$, the function does not have to decay at all as we move away from the cylinder. As we will see it may even diverge. Therefore we have to be very careful about assigning limit values at infinity to functions $\uu \in H^1_w(\Omega)$.



What can be said, based on \cite{girault1991well}, is that one cannot specify boundary conditions simultaneously on the cylinder and at infinity. That is, having specified conditions on the cylinder, the conditions at infinity have already become specified. We will see that the resolution of the Stokes paradox is simple once we know in what function spaces to look for solutions. We now know that solutions exists for the Stokes problem, but only in a certain weighted Sobolev space. Due to the weight going to zero as we move away from the cylinder, the solution is allowed to behave more mischievously in this region. In the time of Stokes (1819--1903), the functional analysis approach to partial differential equations was still in its infancy. Indeed, it was not until 1991 \cite{girault1991well} that the appropriate function spaces were fully clarified.

\subsection{Derivation of the Stokes paradox}
\label{sec:derivation}
Now that we know there exists a solution, we can straightforwardly formulate the Stokes paradox. For this, it is useful to choose moving coordinates. Instead of thinking of a fixed cylinder in a moving fluid, let us reverse the point of view by using moving coordinates such that the fluid appears at rest. If we think of a moving cylinder in an infinite fluid, we can pose the (Navier--)Stokes equations as in \eqref{eqn:firstnavst} with a boundary function $\gbc=(1,0)$, assuming
the cylinder is moving in the $x$-direction with unit speed. In view of \cite{girault1991well}, there is a unique solution $\uu\in H^1_w(\Omega)$, where $\Omega$ is the complement of the cylinder (i.e. $\{  (x,y)\in \mathbb{R}^2: x^2+y^2>1 \}$) and fixed in time.

But $\gbc\in H^1_w(\Omega)$, and $\gbc$ is a solution of \eqref{eqn:firstnavst},  with constant pressure.
And \cite{girault1991well} proves that $\gbc$ is the solution.
Thus we have proved the following theorem.

\begin{theorem}\label{thm:stokesparadox}
Suppose that we move an infinite cylinder in a direction perpendicular to the axis of the cylinder with unit speed. If the entirety of the fluid is governed by the Stokes equations \eqref{eqn:firstnavst} with a no-slip boundary condition on the cylinder, then the entirety of the fluid is forced to move at unit speed.
\end{theorem}

In the variational framework, the Stokes paradox is not really a paradox: The Stokes equation is well posed in infinite domains, but the appropriate function space for $\uu$ is one where we give up control of $\uu$ as it approaches infinity. Thus it is not surprising that the solution is non-physical away from the cylinder.

The fact that we are not able to specify the limiting value of the solution raises the question of how badly the solution might behave. We investigate this in the next section.
\subsubsection{Limit values of functions in $H^1_w$}

In the previous section we saw that moving an infinitely long cylinder through an infinite domain $\Omega$ with given speed $\gbc=(1,0)$ caused the entire solution flux to be $\uu=(1,0)$. In fact,
due to the weight function, any $\vv\in H^1_w(\Omega)$ can tend to a nonzero constant at infinity. Worse, it can grow like $(\log r)^\alpha$ for $\alpha<1/2$, as long as its gradient remains square integrable.
In particular, take $u(r)=(\log r)^\alpha$.
Then for $r>1$,
$$
|\nabla u|= \Big|\alpha \frac{\nabla r}{r} (\log r)^{\alpha-1}\Big|
= \Big|\alpha \frac{\xx}{r^2} (\log r)^{\alpha-1}\Big|
= \Big|\frac{\alpha}{r} (\log r)^{\alpha-1}\Big|.
$$
This expression is square integrable at infinity if
$$
\int_K^\infty \frac{r\,dr}{r^2 (\log r)^{2(1-\alpha)}}<\infty.
$$
Changing coordinates to $s=\log r$ (so that $r^{-1}dr=ds$), our condition reduces to
$$
\int_{\log K}^\infty \frac{ds}{s^{2(1-\alpha)}}<\infty.
$$
This holds when $\alpha<1/2$.

Thus the Stokes equation posed in an infinite domain may have a solution that diverges as $\vert\xx\vert\rightarrow \infty$. 
An example of this is the zero friction solution \eqref{eqn:zerofriction},
as we now discuss.

\section{Navier's revenge}
\label{sec:navrevenge}

In Section \ref{sec:derivation} we saw how the imposition of a no-slip boundary condition on the cylinder leads to the Stokes paradox in unbounded domains. In this section we will discuss what may happen for different boundary conditions. We will see that the Stokes paradox does not occur for all boundary conditions.

Instead of the no-slip boundary condition \eqref{eqn:bceesnavst} with $\mathbf{g}=0$, let us consider Navier's slip condition. This boundary condition, sometimes referred to as Navier's friction condition \cite{ref:SlipNewtonianReviewexpts,lrsBIBiy,dhifaoui2019weighted},
links the tangential velocity and the shear stress on $\Gamma$:
\begin{equation} \label{eqn:navierslip}
\beta \, \uu\cdot\btau_k = -\nu \, \nn^t(\nabla\uu+\nabla\uu^t)\btau_k,\quad k=1,2,
\end{equation}
where $\btau_i$ are orthogonal tangent vectors and $\beta$ is the friction coefficient. This is coupled with the no-penetration condition $\uu\cdot\nn=0$ on $\Gamma$.
In our two-dimensional case of flow around a cylinder, there is only one tangent vector $\btau$. The other one is perpendicular to the plane of the two-dimensional flow, that is, parallel to the cylinder axis. 
For $\beta>0$, the Navier slip condition works as a friction causing the fluid 
to slow down as it slips over the cylinder boundary $\Gamma$.

\omitit{
At this point, it is interesting to ask if exchanging the no-slip boundary condition with Navier's slip condition resolves the Stokes paradox. Before attempting to prove such a result, it is often instructive to look for a counter example. With this in mind, let us consider the two analytic solutions derived in Section \ref{sec:analytic}.
}

The potential function $\pfn$ and stream function $\sfn$ defined in \eqref{eqn:sokesolst} and \eqref{eq:stream-2} give rise to two different solutions of the Stokes equations with Navier's boundary condition \eqref{eqn:navierslip},
each corresponding to a particular choice of $\beta$. 
The potential flow solution \eqref{eqn:yousokesolst}, i.e.
\begin{equation*}  
\begin{split}
\uu=(\pfn_x, \pfn_y)\quad \Rightarrow \quad
\uu(x,y)=\Big(1+\frac{y^2-x^2}{(x^2+y^2)^2},
\frac{-2xy}{(x^2+y^2)^2}\Big),
\end{split}
\end{equation*}
solves the Navier--Stokes equations for all $\nu$, and satisfies the Navier slip condition 
if $\beta=-2\nu$ \cite{lrsBIBiw}. 
Moreover, this flow goes to the desired asymptotic limit (zero) at infinity. 
Thus \eqref{eqn:yousokesolst} resolves Stokes' paradox for $\beta=-2\nu$. 
For this particular boundary condition the solution belongs to the standard Sobolev 
space $H^1(\Omega)$ and exhibits reasonable physical behavior in the entire domain.

This raises the question of what happens for other values of $\beta$.
Interestingly, the other analytic solution we have, i.e.~\eqref{eqn:zerofriction}, 
satisfies the friction boundary condition \eqref{eqn:navierslip} if $\beta=0$. 
To see this, note that $\btau=(-y, x)$ and $\nn=(-x,-y)$ on $\Gamma$. 
Computing  $(\nabla \uu )\, \btau$ we find
\begin{equation}\label{eqn:treseetermyou}
\begin{split}
(\nabla\uu)\btau &= \btau\cdot\nabla\uu 
= \partial_\theta\big(\sin^2\theta + \log(r), -\cos\theta\sin\theta\big) \\
&= \big(2\sin\theta\cos\theta , -\cos^2\theta+\sin^2\theta\big).
\end{split}
\end{equation}
Therefore (omitting some trigonometric simplifications)
\begin{equation}\label{eqn:nowennmyou}
\begin{split}
\nn^t(\nabla\uu)\btau |_\Gamma
&=-(\cos\theta,\sin\theta)^t \big(2\sin\theta\cos\theta , -\cos^2\theta+\sin^2\theta\big)\\
&=-2\sin\theta\cos^2\theta +\cos^2\theta\sin\theta-\sin^3\theta
=-\sin\theta.
\end{split}
\end{equation}
Similarly
\begin{equation}\label{eqn:enntermyou}
(\nabla\uu)\nn =\nn\cdot\nabla\uu
= -r \partial_r\big(\sin^2\theta + \log(r), -\cos\theta\sin\theta\big)
= \big(-1, 0\big).
\end{equation}
This says that
\begin{equation}\label{eqn:dotaurmyou}
\btau^t(\nabla\uu)\nn|_\Gamma
=(-\sin\theta,\cos\theta)\cdot \big(-1, 0\big)=\sin\theta.
\end{equation}
Note that $\nn^t(\nabla\uu^t)\btau=\btau^t(\nabla\uu)\nn$.
Therefore
\begin{equation} \label{eqn:notquickial}
\nn^t(\nabla\uu+\nabla\uu^t)\btau|_\Gamma=
\big(\nn^t(\nabla\uu)\btau+\btau^t(\nabla\uu)\nn\big)|_\Gamma=0.
\end{equation}

\omitit{
\begin{align}\label{eqn:gradutau}
    (\nabla \uu)\, \btau = \frac{y^2-x^2}{r^4} \btau \quad \Rightarrow \quad \nn^t (\nabla \uu) \btau=\frac{y^2-x^2}{r^4} \nn \cdot \btau = 0 .
\end{align}
Next, we compute $\nn^t (\nabla \uu)^t \btau$. One way is to take the transpose of this expression (which is equivalent as it is a scalar), and use that $(\nn^t (\nabla \uu)^t \btau)^t=\btau^t (\nabla \uu) \nn$.
Again, we can compute
\begin{align}\label{eqn:gradun}
    (\nabla \uu) \nn = \frac{y^2-x^2}{r^4} \nn,
\end{align}
meaning that $\nn^t (\nabla \uu)^t \btau=0$.
}

Thus the function in \eqref{eqn:zerofriction}
solves the Stokes equations and satisfies the Navier slip condition if $\beta=0$.
But this solution diverges as $r\rightarrow \infty$, fast enough that 
the norm in \eqref{eqn:dubyahonet} is not finite.
Thus \eqref{eqn:zerofriction} does not resolve the Stokes paradox.

It is worth noting that the fact that $\beta=0$ does not mean that the drag
on the cylinder is zero \cite{lrsBIBjn}.
It is known that the drag IS zero for $\beta=-2\nu$, and this is the core
of d'Alembert's paradox \cite{lrsBIBjn}.

The computations above are sufficiently complex that it is useful to have a way to verify them. This can be done by solving the Stokes equations with the Navier slip condition numerically with $\beta=0$ and check that the result approximately agrees with \eqref{eqn:zerofriction}.

For $\beta>0$, the Navier slip condition acts as a friction force, slowing the flow as it slips over the cylinder. 
For $\beta\to\infty$, the Navier friction boundary condition converges to the Stokes no-slip condition. 
For other values of $\beta$, the techniques in section \ref{sec:bounded-domains}
could be used to see if there are plausible solutions of the Stokes paradox.

In conclusion, we see that the Stokes paradox is resolved using the Navier slip boundary condition with one particular value for $\beta$, but not for others.
Navier died in 1836, so he was not available to comment on Stokes' paradox. We can only wonder what he might have said.

\section{Stokes flow on bounded domains of increasing size}
\label{sec:bounded-domains}
In Section \ref{sec:sobolev} we saw how the Stokes problem lost the ability to specify the value of the solution on the boundary away from the cylinder. With this in mind, we now restrict our attention to bounded domains, where it is possible to pose boundary conditions. We explore two approaches, a computational one and an analytical one. We will see that as we increase the size of the box, we again encounter the Stokes paradox.

\subsection{Computational approach}
\label{sec:compapp}

Recent advances in software \cite{alnaes2015fenics,hecht2012new} have made it easy to solve partial differential equations (PDEs). Using such software, you can study PDEs without knowing detailed background prerequisites \cite{lrsBIBih}.
We now indicate this approach for the Navier--Stokes equations.

Consider the domain $\Omega_b$ defined by
\begin{equation} \label{eqn:ohmercst}
\Omega_b=\set{\xx}{|\xx|>1,\; |x_i|<b,\;i=1,2}
\end{equation}
for $b>1$. Let $\Gamma$ denote the subset of $\partial\Omega_b$ defined by
$$
\Gamma=\set{\xx}{|\xx|=1},
$$
that is, $\Gamma$ represents the cylinder.

\begin{figure}
\centerline{\includegraphics[width=2.5in,angle=0]{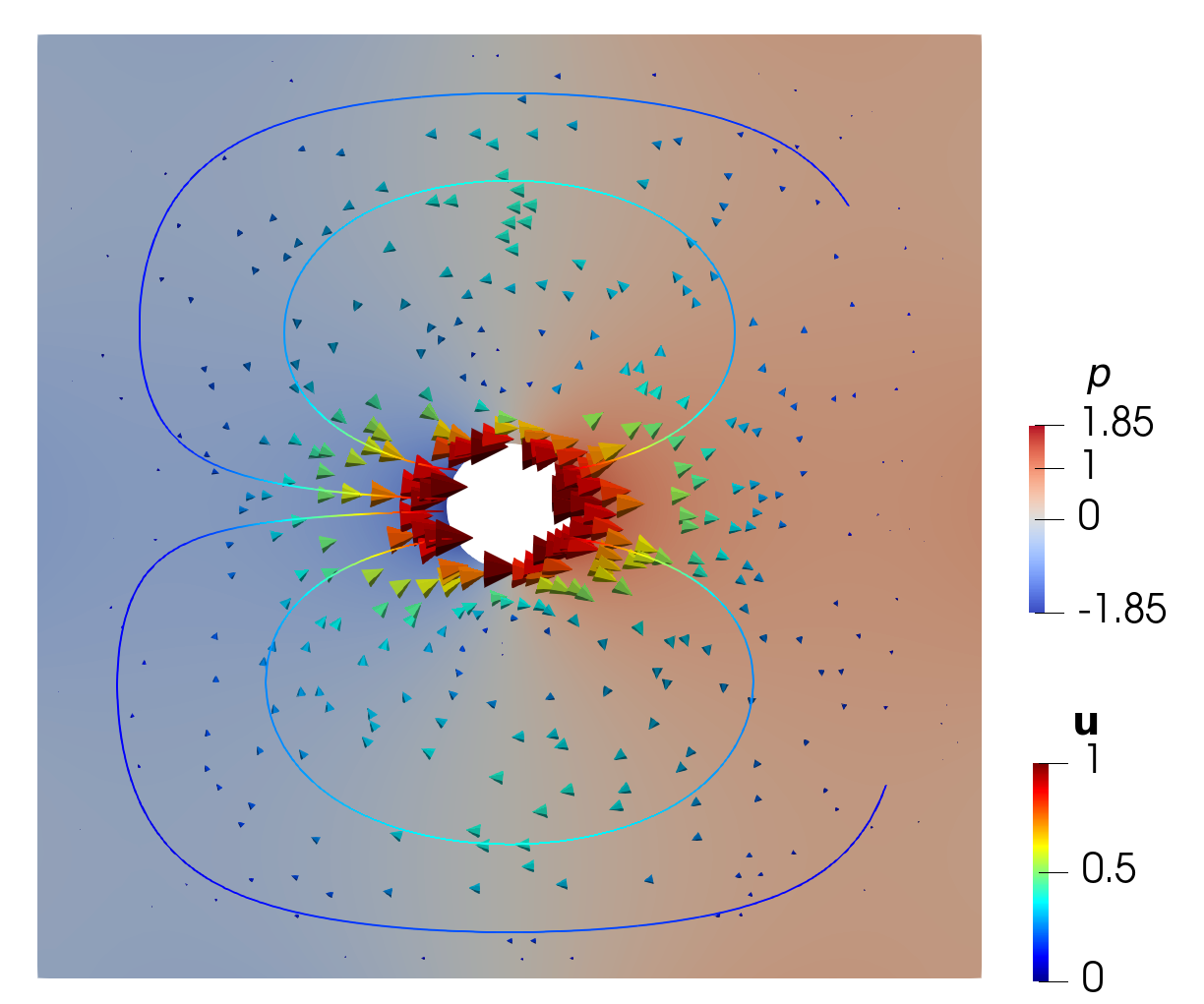}}
\vspace{-0.0cm}
\caption{Plot of pressure $p$, flux $\uu$ (glyphs) and streamlines for the solution the moving cylinder problem \eqref{eqn:firstnavst} in the domain \eqref{eqn:ohmercst} with boundary conditions \eqref{eqn:specbcsst} for $b=7.5$. Due to no-slip boundary condition $\uu=(0,0)$ on the box walls, the fluid is forced to recirculate.
}
\label{fig:paravec}
\end{figure}

We keep our viewpoint of a cylinder moving through the larger domain where the fluid is
at rest.
I.e., we consider solutions $\uu^b$ of the problem \eqref{eqn:firstnavst} with boundary conditions
\begin{equation} \label{eqn:specbcsst}
\uu^b=(1,0)\;\hbox{on}\;\Gamma,\qquad 
\uu^b=\bfz\;\hbox{on}\;\partial\Omega_b\backslash\Gamma.
\end{equation}
Figure \ref{fig:paravec} shows the solution for $b=4$.

\begin{figure}
\centerline{(a)\includegraphics[width=1.6in,angle=0]{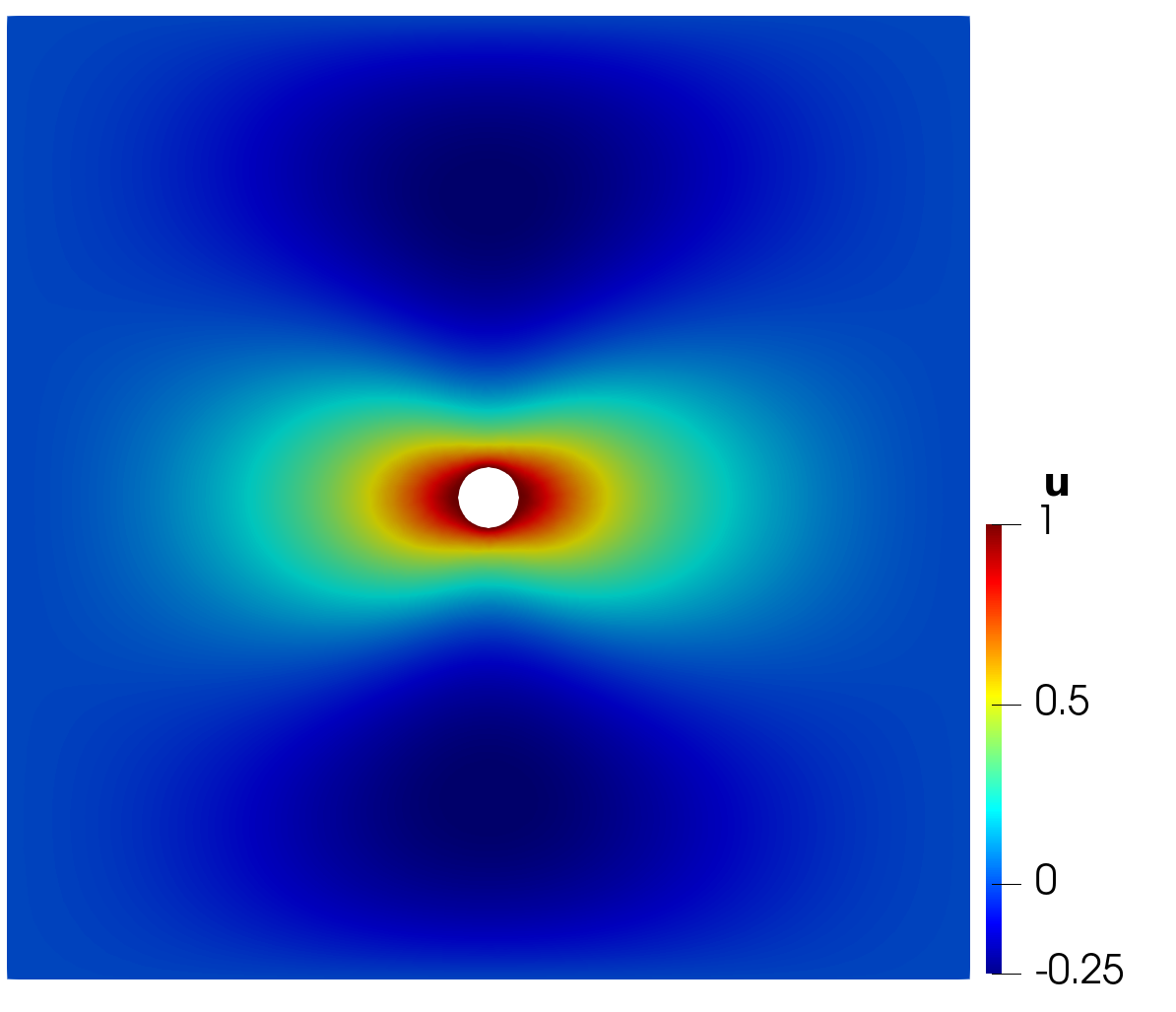}\quad
(b) \includegraphics[width=1.6in,angle=0]{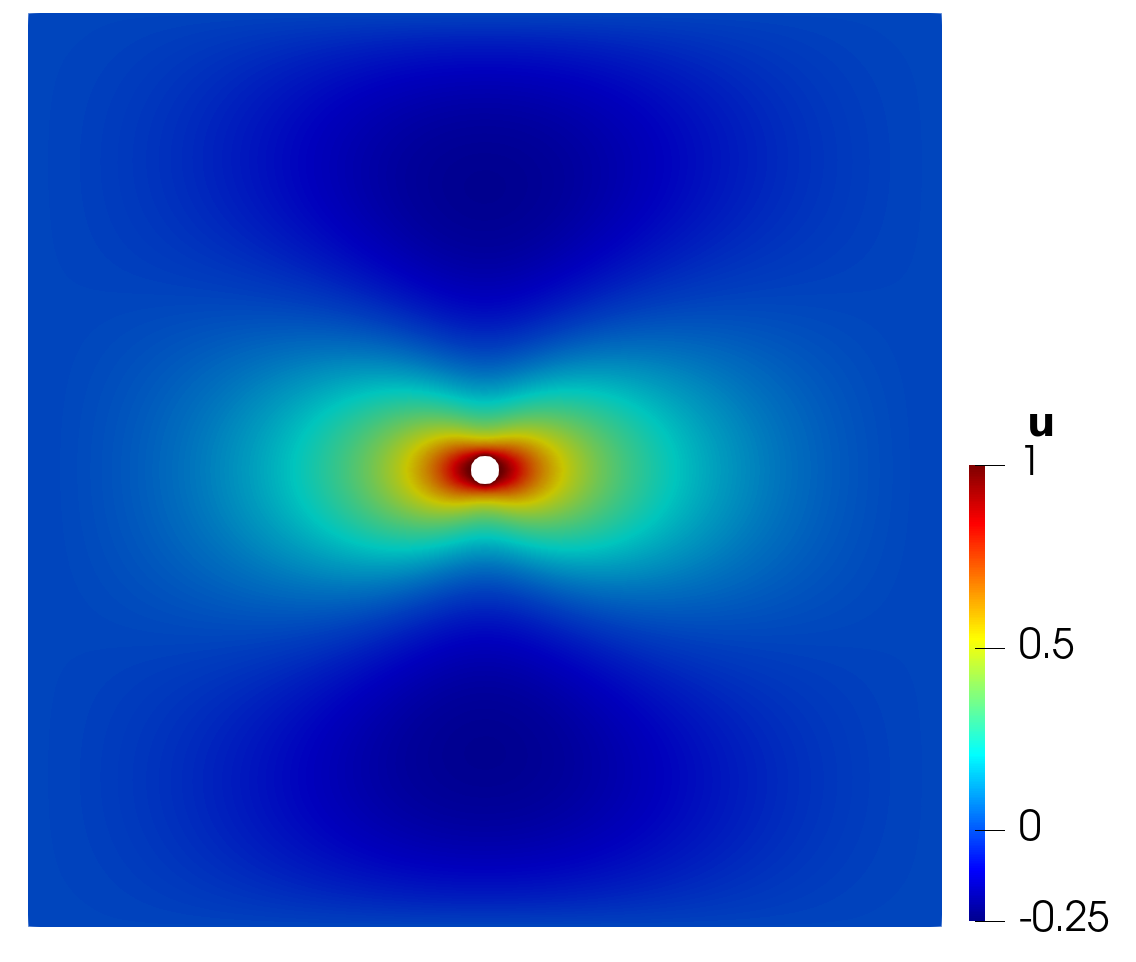}
(c) \includegraphics[width=1.6in,angle=0]{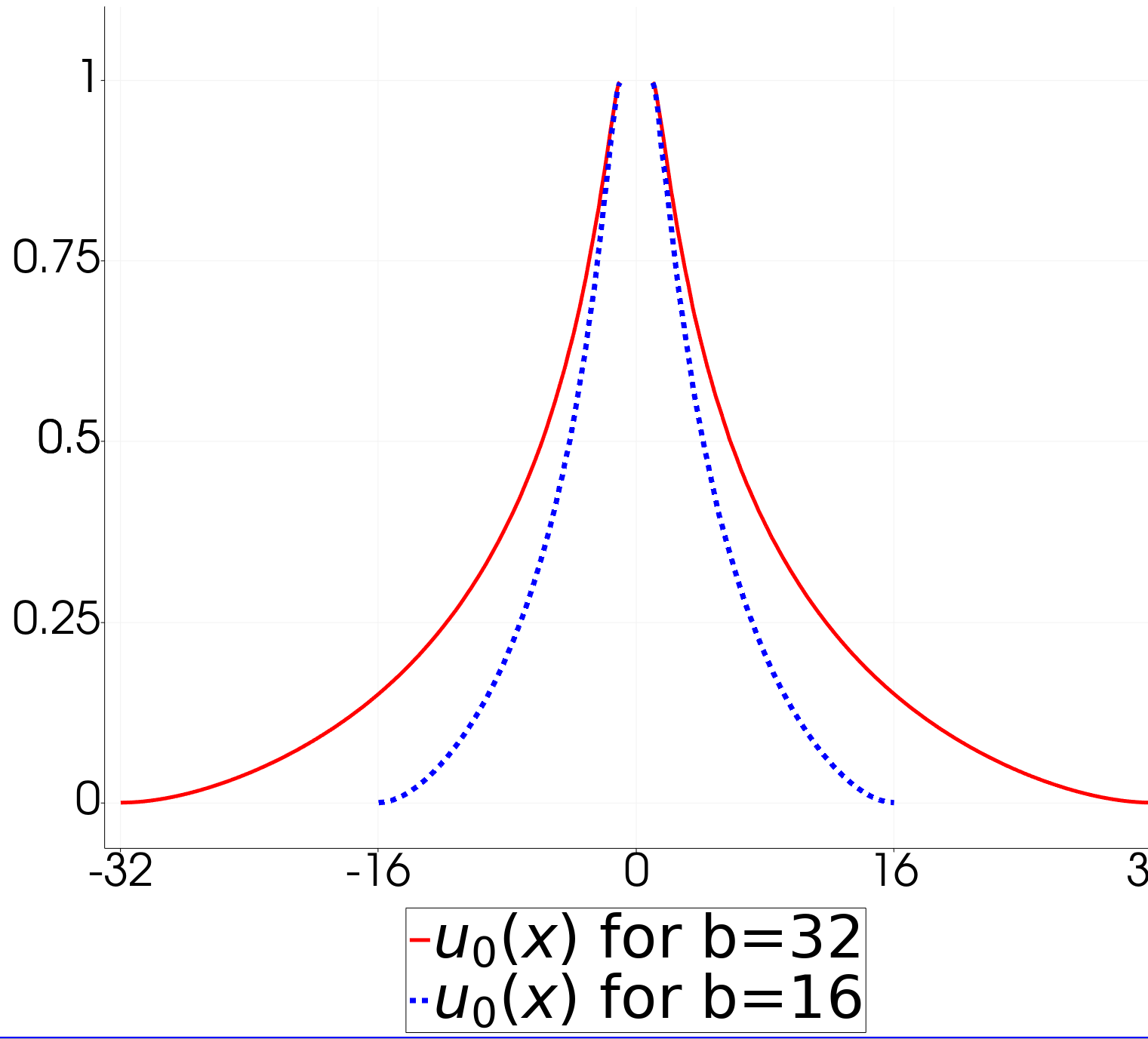}}
\vspace{-0.0cm}
\caption{Plot of the horizontal component of the solution of \eqref{eqn:firstnavst}
in the domain \eqref{eqn:ohmercst} with boundary conditions \eqref{eqn:specbcsst} for
(a) $b=16$, $M$=64 and (b) $b=32$, $M$=128. $M$ is the mesh parameter for {\tt mshr}, with the number of segments for the definition
of the circle chosen to be M as well.}
\label{fig:paradokes}
\end{figure}

Figure \ref{fig:paradokes} shows the horizontal component of the
solution for (a) $b=16$ and (b) $b=32$. We see that the support of the horizontal component spreads as the box gets bigger. Thus we see that the horizontal component of the solutions is not really going to zero at the boundary of the box. It remains positive as we go to the edge of the domain both upstream and downstream of the cylinder.

To examine how the support of the horizontal component of the solution spreads
as $b$ is increased, we considered a functional to examine the size of $\uu^b$
in regions of increasing size $d$, but fixed independent of $b$.
Thus we defined
\begin{equation} \label{eqn:growfunl}
\int_{\Omega_b} \chi_d(\xx)^2 |\uu^b(\xx)|^2\,d\xx\Big/
\int_{\Omega_b} \chi_d(\xx)^2 \,d\xx
\end{equation}
where $\chi_d(\xx)$ is the interpolant on the computational mesh of the cut-off function
$$
\frac12\Big(1-\tanh\big(20\big(|\xx|^2-d^2\big)\big)\Big)
$$
which is very close to 1 inside $|\xx|<d$ and very close to zero outside of that. If $\uu^b\to(1,0)$ as $b\to\infty$, then we would expect the expression \eqref{eqn:growfunl}
to increase to 1. If on the other hand, if $\uu^b\to\bfz$ as $r\to\infty$, we would expect the
expression \eqref{eqn:growfunl} to converge to some value less than 1 as $b$ is increased.

\begin{figure}[h]
\centerline{(a)\includegraphics[width=2.4in,angle=0]{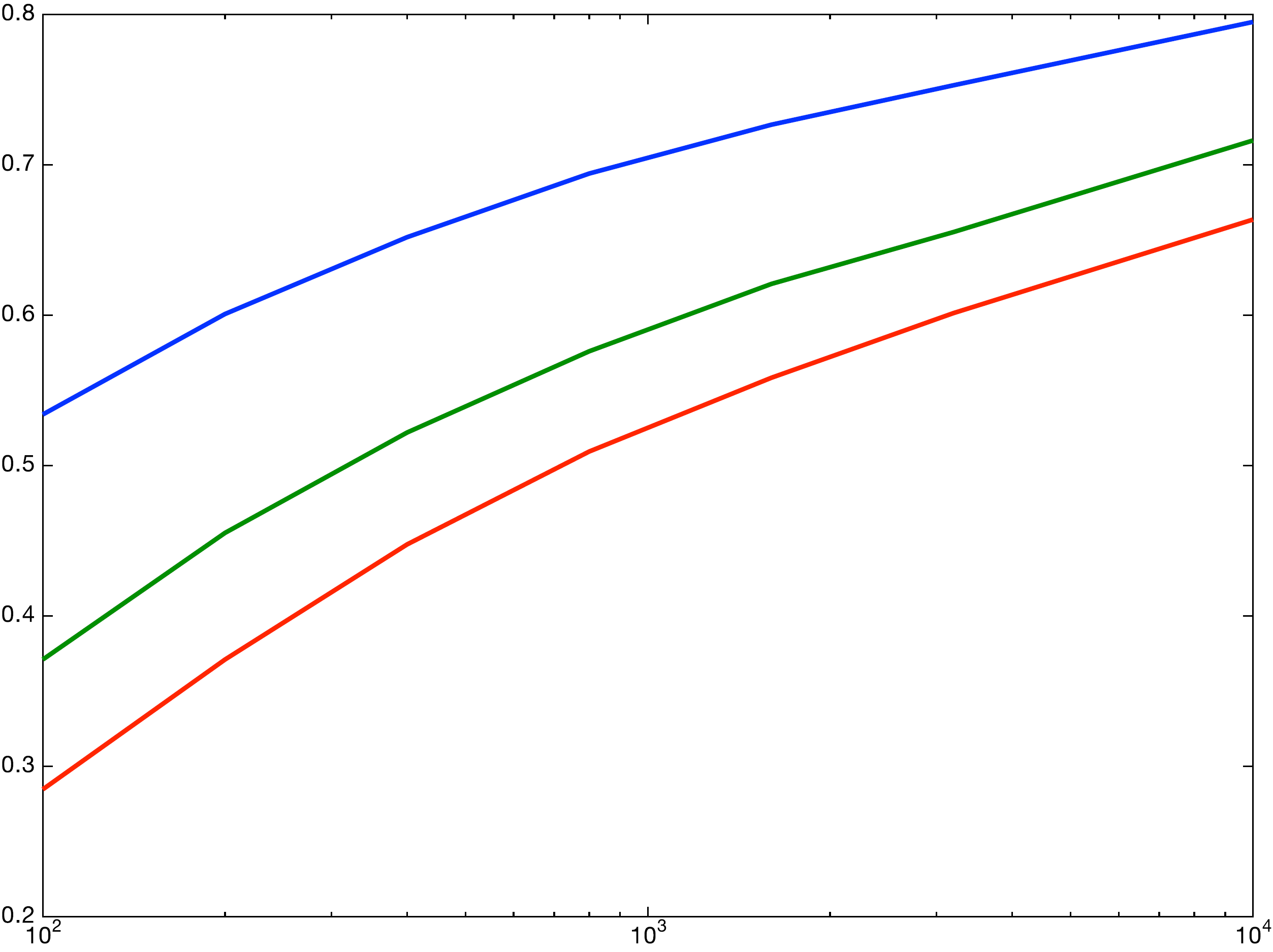}\quad
            (b)\includegraphics[width=2.4in,angle=0]{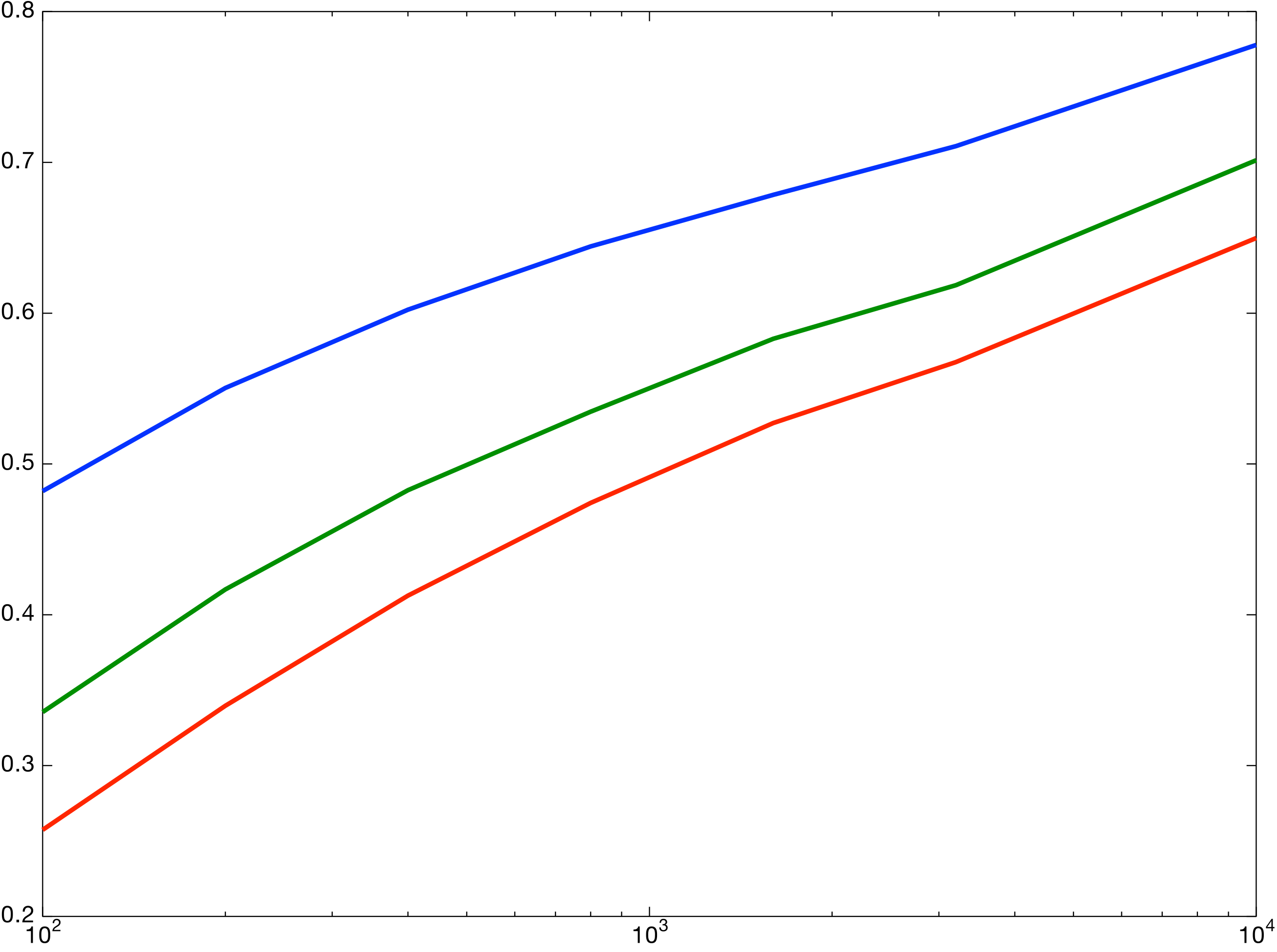}}
\vspace{-0.0cm}
\caption{Growth of \eqref{eqn:growfunl} as a function of $r$ (horizontal axis)
for three values of $d$: (top) $d=10$, (middle) $d=20$, (bottom) $d=30$.
(a) Stokes no-slip boundary condition, (b) Navier friction boundary condition, $\beta=0$.}
\label{fig:tripara}
\end{figure}

Figure \ref{fig:tripara} gives the data for three values of $d$ as a function of box size $b$. It appears \eqref{eqn:growfunl} indeed increases to 1, which points to $\uu^b\to(1,0)$ for $\vert \xx \vert < d$ as $b\to\infty$. This is in accordance with the Stokes paradox as stated in Theorem \ref{thm:stokesparadox}; that as $b \rightarrow \infty$, we have $\uu=(1,0)$ everywhere. But then the fluid moves like a solid.   

Interestingly, $\uu=(1,0)$ satisfies the Navier slip condition with $\beta=0$ since $\nabla \uu=\bfz$. Thus this solution not only satisfies the no-slip boundary condition on the cylinder, but also the Navier friction condition with $\beta=0$. The other solution with $\beta=0$, i.e. \eqref{eqn:zerofriction} has different boundary values on the cylinder. It also diverges when $r \rightarrow \infty$, unlike the solution $\uu=(1,0)$.

\subsection{Analytic solutions in bounded domains}
\label{sec:analytic}
Let us return from modern numerical software back to the classics.
In this section, we consider analytical solutions in increasingly large circular domains,
following \cite{shaw2009simple}.
These domains are related to the so-called Leray approximate solutions \cite{amick1988leray}.

\omitit{
Recall the function $\sfn(r,\theta)=f(r) \sin\theta$ where $f(r)=r\log r$.
Using polar coordinates, we have
\begin{equation} \label{eqn:biharmnl}
\begin{split}
\Delta \sfn&= \sin\theta\Big(f''(r)+r^{-1} f'(r)-r^{-2} f(r)\Big)\\
&= \sin\theta\big(r^{-1} + r^{-1}(1+\log r) -r^{-1} \log r\big)
=\frac{2y}{r^2},
\end{split}
\end{equation}
where we used $f'(r)=1+\log r$ and $f''(r)=r^{-1}$.
Similar to the function in \eqref{eqn:sokesolst},
we have  $\Delta (2y/r^2)=0$.
Thus $\sfn$ is biharmonic, i.e. $\Delta^2 \sfn=0$,
in any open set that excludes the origin.
}

Following \cite[(12)]{ref:tannerpowerlawparadox},
consider a general biharmonic stream function of the form
$$
\psi=f(r) \sin\theta,\qquad f(r)=Ar^{-1}+Br\log r + C r^3+Dr.
$$

Now let us show that the fact that $\psi$ is biharmonic implies that
$\uu=(\psi_y,-\psi_x)$ satisfies the first equation in \eqref{eqn:firstnavst}. For the sake of calculations, let us for the moment augment the domain with a $z$-component and let $\boldsymbol{\psi}=(0,0,\psi)$ so that we can define $\uu=\curl \boldsymbol{\psi}$.
Note that $\nabla \cdot\boldsymbol{\psi}=0$ since $\psi$ depends only on $x$ and $y$.
By using the vector calculus identity
$\curl(\curl \vv)=\nabla(\nabla \cdot \vv) - \Delta \vv$,
we then see
\begin{align*}
\curl\Delta\uu &=\curl (\Delta (\curl \psi)) = \curl \curl \Delta \boldsymbol{\psi} =\nabla \underbrace{(\nabla\cdot \Delta \boldsymbol{\psi})}_{=0}-\Delta^2 \boldsymbol{\psi}.   
\end{align*}
Since all four terms in $\psi$ are biharmonic in any open set that excludes the origin, we can then conclude that $\uu$ satisfies the following: $\curl\Delta\uu=-\Delta^2 \boldsymbol{\psi}=0$ in any open set that excludes the origin.

Invoking Stokes' theorem \cite[Theorem 2.9]{giraultraviart}, we conclude
that $\Delta\uu=\nabla p$ for some scalar function $p$.
Thus $\uu$ satisfies \eqref{eqn:firstnavst}.
Since $\uu$ has the $z$-component zero, $p_z=0$, and hence $p$ is constant
in $z$.
Subtracting this constant, we can view $p$ as being zero in the $z$-component
and in this sense independent of $z$.
So we have proved that $\uu$ is a solution of the two-dimensional Stokes equations.

\omitit{
Now, consider the following change-of-coordinate expressions. Differentiating
\begin{align*}
    r=\sqrt{x^2+y^2}
\end{align*}
we find
\begin{equation}\label{eqn:chacoronlybis}
r_x=\frac{x}{r},\qquad r_y=\frac{y}{r}.
\end{equation}
Similarly
\begin{equation}\label{eqn:chacorarain}
\theta_x=\frac{-x\sin\theta}{r^2\cos\theta}
=\frac{-\sin\theta}{r},\qquad
\theta_y=\frac{y\cos\theta}{r^2\sin\theta}
=\frac{\cos\theta}{r}.
\end{equation}
These imply that
\begin{equation}\label{eqn:xxxconsrain}
\begin{split}
-u_y=\psi_x&=f'r_x\sin\theta+(f\cos\theta)\theta_x
=\frac{xf'\sin\theta}{r}-\frac{xf\sin\theta}{r^2}\\
&=x\sin\theta\bigg(\frac{f'}{r}-\frac{f}{r^2}\bigg).
\end{split}
\end{equation}
The quantity $u_i$ indicates the $i$th component of the vector $\uu$.
Similarly, recalling that $y=r\sin\theta$, we have
}
Using polar coordinates, we find
\begin{equation}\label{eqn:sighmorn}
\begin{split}
-u_y
=x\sin\theta\bigg(\frac{f'}{r}-\frac{f}{r^2}\bigg), \qquad
u_x
={f'\sin^2\theta}+\frac{f\cos^2\theta}{r}.
\end{split}
\end{equation}
Impose constraints
\begin{equation}\label{eqn:consrain}
f(1)=f'(1)=1,\quad f(b)=f'(b)=0.
\end{equation}
The latter two constraints in \eqref{eqn:consrain} imply that $\uu=\curl\psi=\bfz$ for $r=b$.
The first two constraints in \eqref{eqn:consrain} imply that
$$
\uu(r=1)=(1,0).
$$
Since we have identified four parameters and four constraints, we likely have
found the required solution.
But to be sure, we need to solve these equations and see what happens when $b\to\infty$.

\subsubsection{Algebraic solution of the PDE}
\label{sec:alg-pdes}
Using the boundary conditions \eqref{eqn:consrain},
 we can evaluate the constants $A$, $B$, $C$, and $D$.
We have
$$
B=\frac{-2(b^2+1)}{2 + 2b^2(\log b -1)+2\log b}
 =\frac{-(1+b^{-2})}{ \log b -1+b^{-2}(1+\log b)}
\approx\frac{-1}{\log b}
$$
and
$$
C=\frac{1}{2 + 2b^2(\log b -1)+2\log b} \approx\frac{1}{2b^2\log b},
$$
together with
$$
A=\half B +C\qquad\hbox{and}\qquad D=1-\half B -2 C.
$$
Although its derivation is tedious and error-prone, such a result
can be checked in various ways.
Thus as $b\to\infty$,
$$
B\to 0,\;b^2 C\to 0\implies A\to 0,\; D\to 1.
$$
Therefore $\uu^b\to (1,0)$ as $b\to\infty$.

\omitit{
$$
 f(r)\approx r - \frac{r\log r}{\log b} + \frac{r^3}{2b^2\log b}-\frac{1}{2r\log b}
$$
}

\subsubsection{Asymptotic behavior}

\begin{figure}[h]
\centerline{\includegraphics[width=4.6in,angle=0]{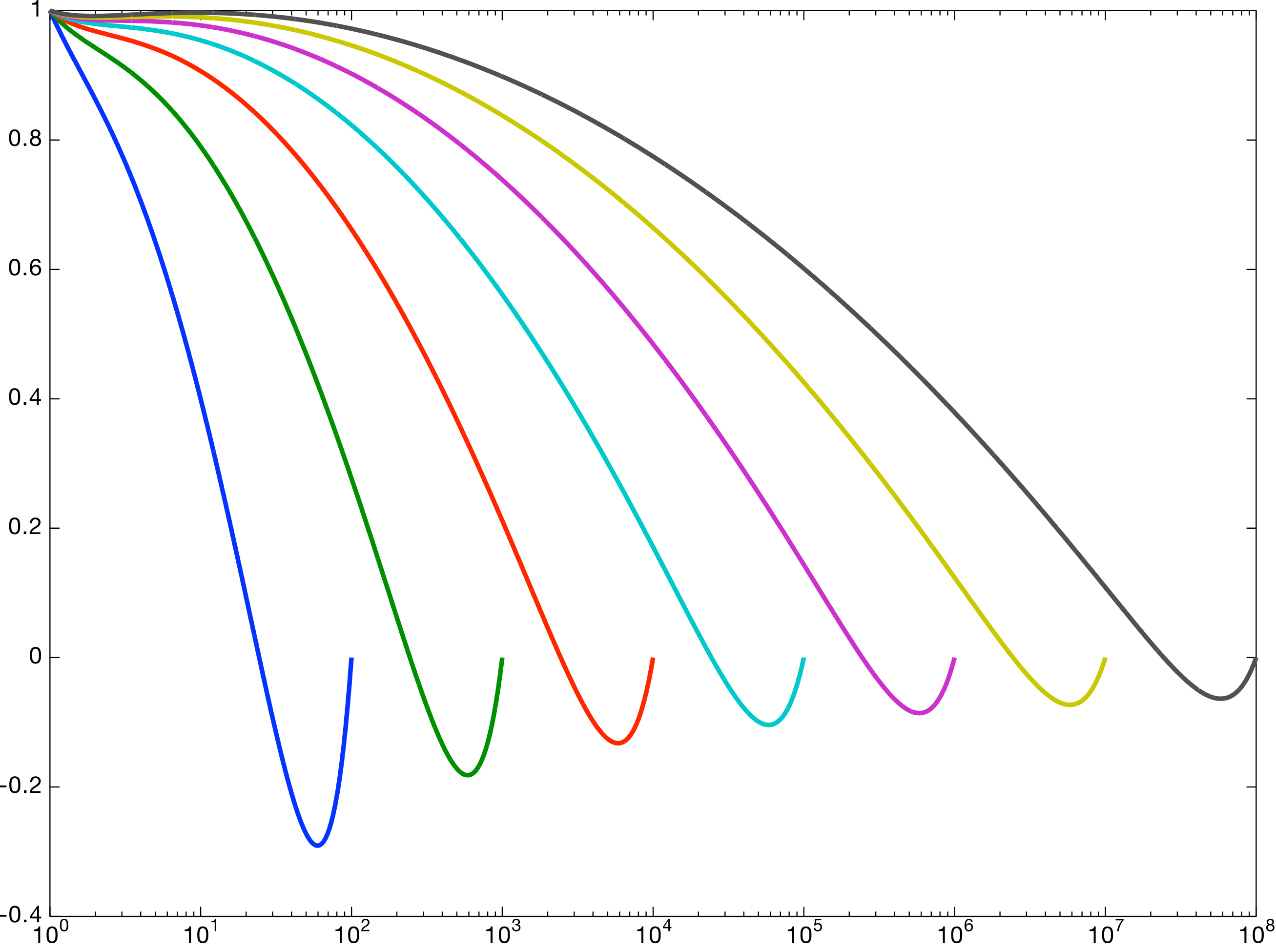}} 
\vspace{-0.0cm}
\caption{Plot of $\chi_b$ defined in \eqref{eqn:fideclin} for $b=10^k$ for
$k=2,3,\dots,8$.
The horizontal axis is $r$.
}
\label{fig:ploteffpee}
\end{figure}

In particular, $A$, $B$, and $b^2C$ decay like $1/\log b$.
Using 
\eqref{eqn:sighmorn}, we find
$$
\uu(r,\theta)=(f'(r),0)-\big(f'(r)-r^{-1}f(r)\big)(\cos^2\theta,\cos\theta\sin\theta).
$$
Subtracting the expressions for $f'$ and $f/r$, we find
$$
\big|f'(r)-r^{-1}f(r)\big|=\bigg|\frac{-2A}{r^2}+B+2Cr^2\bigg|\leq \frac{c}{\log b}.
$$
Examining the expression for $f'$, we see that it decays like $1/\log r$.
Thus we considered the expression
\begin{equation}\label{eqn:fideclin}
\chi_b(r)=\Big(1+\frac{3\log r}{2\log b}\Big) f'(r).
\end{equation}
A plot of $\chi_b$ for $b=10^k$ for $k=2,3,\dots,8$
is seen in Figure \ref{fig:ploteffpee}.
From this figure, we see that $\chi_b\approx 1$ for small $r/b$.
Note that, by definition of $\chi_b$,
\begin{equation}\label{eqn:defeclin}
u_x\approx f'(r)=\Big(1+\frac{3\log r}{2\log b}\Big)^{-1} \chi_b(r).
\end{equation}

\subsection{Friction boundary conditions}

We performed a series of tests solving \eqref{eqn:firstnavst} in the
domain \eqref{eqn:ohmercst} with Navier boundary conditions \eqref{eqn:navierslip}
with $\beta=0$, for  various $r$. \omitit{Shown in Figure \ref{fig:paranavier} is the case $r=32$.
We see that, except near the cylinder, the flow pattern is similar to the
case of Stokes no-slip boundary conditions ($\beta=\infty$).} Figure \ref{fig:tripara}(b) gives the data for three values of $d$ as a function of
box size $r$ for Navier boundary conditions \eqref{eqn:navierslip} with $\beta=0$.
These data suggest that $\uu_r$ is converging to $(1,0)$ as $r\to\infty$
with Navier boundary conditions.

\omitit{
In Figure \ref{fig:paradiff}, we show the difference between 1 and
the horizontal component of the solution, of \eqref{eqn:firstnavst} in the
domain \eqref{eqn:ohmercst} with no-slip boundary conditions,  for $r=33$.
What is not clear is why the error is completely uniform.
At the boundary, where we impose zero boundary conditions explicitly in {\tt dolfin},
the error should be of order 1.
Results for larger values of $r$ are similar.
The code {\tt dolfin} that was used for the computations effectively uses a
penalty method to enforce Dirichlet boundary conditions, so this could be the
cause of the discrepancy.

\begin{figure}
\centerline{\includegraphics[width=6.6in,angle=0]{paradokesM128r33.png}}
\vspace{-0.0cm}
\caption{Plot of the difference between the horizontal component of the solution
(of \eqref{eqn:firstnavst} in the domain \eqref{eqn:ohmercst} with no-slip boundary conditions)
and 1, for $r=33$.
The mesh parameter M=128 for {\tt mshr}, with the number of segments for the definition
of the circle chosen to be M as well.}
\label{fig:paradiff}
\end{figure}
}
\section{Navier--Stokes: no paradox}
\label{sec:nsnopar}

According to \cite[corollary to Theorem 7A]{ref:FinnExterioreview}, the nonlinear
problem \eqref{eqn:effnavstot} has a solution with
$\gbc=\bfz$ and $\uu\to \uu^\infty$ with $\uu^\infty$ a constant, provided that
$|\uu^\infty|$ is sufficiently small;
also see \cite[Theorem XII.5.1]{ref:Galdi2011IntroMathofNS}.
The realization that adding an advection term to the equations resolves
the Stokes paradox began with the work of Oseen \cite{oseen1927neuere}.
See \cite{ref:FinnExterioreview} for more historical references.
The results of Finn \cite{ref:FinnExterioreview} confirm that, for the Navier--Stokes
equations, one can pose boundary conditions both on the cylinder (or other bluff body)
and at infinity.

The constant function $\gbc$ is also a solution
of \eqref{eqn:effnavstot} (with constant pressure) for any $R>0$.
But the boundary conditions are different in this case.
We can sum up the Stokes paradox by saying that a boundary condition is lost
when we set the Reynolds number $R$ to zero.
Thus fluid flow can be described accurately in unbounded domains only by a nonlinear system.

For the cylinder problem, the diameter $L$ gives us a length scale.
Once we pick the flow $\uu^\infty$ (or $\gbc$), we have a speed $U$, and together with the
kinematic viscosity $\nu$, this determines a Reynolds number $R>0$ given by
\eqref{eqn:renodefvst}.
The only way $R$ can be zero is to have $\uu^\infty=\gbc=\bfz$ (or infinite viscosity,
which does not sound like a fluid).
Thus the Stokes equations can be viewed as an approximation for small Reynolds numbers,
and this approximation works well for bounded domains.
But it fails for infinite domains.

The existence of solutions of the Navier--Stokes system for large external flows,
or equivalently for large Reynolds numbers, is reviewed by Galdi in
\cite[section XII.6]{ref:Galdi2011IntroMathofNS}.
However, the results there are not definitive;
they present a condition that must hold if no such solutions exist.
\omitit{Todo: Bring this back? Thus there is a gap in our knowledge about high-speed flows past a cylinder.
Figure \ref{fig:baseflow} gives evidence for such flows.
Other computations of this type \cite{lrsBIBiw,lrsBIBjn} indicate that external
flows can be effectively computed for large Reynolds numbers.}

\section{Extensions of the Stokes paradox}
\label{sec:extox}

The Stokes paradox has implications for other flow problems.
Here we mention two of them.

\subsection{Flow instability}

Determining the form of Reynolds--Orr instability modes for Navier--Stokes flow around a cylinder
requires solution of a generalized eigenproblem of the form \cite{lrsBIBiw}
\begin{equation} \label{eqn:reyorinstt}
\begin{split}
-\Delta \uu + \nabla p &= \lambda^{-1} B_R \uu  \;\hbox{in}\;\Omega,\\
\sdiv\uu &=0\;\hbox{in}\;\Omega,
\end{split}
\end{equation}
with homogeneous boundary conditions on $\Gamma=\partial\Omega$.
Here the multiplication operator $B_R$ is defined by
$$
B_R(\xx)=\half\big(\nabla\uu_R(\xx)+\nabla\uu_R^t(\xx)\big),
$$
where $\uu_R$ solves \eqref{eqn:effnavstot}.
Restricted to a bounded domain, this constitutes a symmetric generalized eigenproblem,
and thus it has real eigenvalues \cite{stewart2001matrix}.

On an unbounded domain, we expect that some rate of decay for $B$ would be required
in order that the eigenproblem is well behaved.
Define 
$$
V=\set{\vv\in H^1_w(\Omega)}{\vv=0\;\hbox{on}\;\Gamma},
$$
and we endow $V$ with the norm of $H^1_w(\Omega)$.

\begin{lemma}\label{lem:beebded}
Suppose that there is a positive constant $C_B$ such that
\begin{equation} \label{eqn:beebded}
|B(\xx)|\leq C_B\big(1+|\xx|^{-2}\log^2|\xx|\big)\quad\forall\xx\in\Omega.
\end{equation}
Then the multiplication operator associated with $B$ is a bounded operator from $V$ to $V'$.
\end{lemma}

In the statement of the lemma, $|B(\xx)|$ denotes the Frobenius norm of $B(\xx)$.
To prove the lemma, recall from \cite[page 315]{girault1991well} that
\begin{equation} \label{eqn:dualnor}
\norm{\uu}_{V'}=\sup_{\bfz\not=\vv\in V}
\frac{\int_\Omega \uu(\xx)\cdot\vv(\xx)\,d\xx}{\norm{\vv}_{H^1_w(\Omega)}}.
\end{equation}
But H{\"o}lder's inequality and \eqref{eqn:beebded} imply
\begin{equation} \label{eqn:holdded}
\begin{split}
\Big|\int_\Omega B(\xx)\uu(\xx)\cdot\vv(\xx)\,d\xx\Big|^2&\leq
\int_\Omega |B(\xx)|\,|\uu(\xx)|^2\,d\xx \int_\Omega |B(\xx)|\,|\vv(\xx)|^2\,d\xx\\
&\leq C_B^2 \norm{\uu}_{H^1_w(\Omega)}^2 \norm{\vv}_{H^1_w(\Omega)}^2.
\end{split}
\end{equation}
Thus we conclude that
$$
\norm{B\uu}_{V'}\leq C_B \norm{\uu}_{H^1_w(\Omega)}.
$$
This completes the proof of Lemma \ref{lem:beebded}.

Consider the operator $K$ defined by $K\vv=\uu$ where $\uu\in V$ solves
\begin{equation} \label{eqn:kayoped}
\begin{split}
-\Delta \uu + \nabla p &= B \vv  \;\hbox{in}\;\Omega,\\
\sdiv\uu &=0\;\hbox{in}\;\Omega.
\end{split}
\end{equation}
Note that the eigenproblem for $K$, that is $K\uu=\lambda\uu$, 
provides a resolution of \eqref{eqn:reyorinstt}.
The following is a corollary of Lemma \ref{lem:beebded}.

\begin{theorem}\label{thm:beebded}
Suppose that \eqref{eqn:beebded} holds.
Then $K$ is a bounded operator from $V$ to $V$.
\end{theorem}

The proof of Theorem \ref{thm:beebded} follows from \cite[Theorem 3.4]{girault1991well} 
and Lemma \ref{lem:beebded}.

From \cite[Remark XII.8.3]{ref:Galdi2011IntroMathofNS} we expect that
$$
|\nabla\uu_R(\xx)|=\order{|\xx|^{-1}\log^2|\xx|}\quad\hbox{for large}\;|\xx|.
$$
Thus  \eqref{eqn:beebded} does not hold for $B_R$, and the associated multiplication
operator is not a bounded operator on $H^1_w(\Omega)$.
Indeed it was found in
\cite{lrsBIBiw} that the eigenvalues increase as the computational domain size
is increased.
We can summarize these observations as follows.
Despite the fact that the Navier--Stokes equations are well defined on 
unbounded domains, the equations for their instabilities are not.
We are tempted to call this the instability paradox.

\subsection{Power-law fluids}

Tanner \cite{ref:tannerpowerlawparadox} has shown that shear thinning power-law
fluids do not suffer Stokes' paradox, but that shear thickening power-law fluids do.
The Stokes power law model is given by \cite[(1.5)]{ref:powerLawStokesPenalty}
\begin{equation} \label{eqn:stopewetot}
\begin{split}
-\nu\sdiv\big(|D\uu|^{r-2}D\uu\big) + \nabla p &=\ff \quad \hbox{in}\;\Omega,\\
\sdiv\uu &=0 \quad \hbox{in}\;\Omega, \\
\end{split}
\end{equation}
where $D\uu= \half\big(\nabla\uu+\nabla\uu^t\big)$.
The fluid model is shear thinning if $r<2$ and shear thickening if $r>2$.
The case $r=2$ is the standard Stokes model.

Tanner showed that for flow around a cylinder, the Stokes paradox holds for $r>2$,
but not for $r<2$.
The approach \cite{girault1991well} can possibly extend this result to
more general domains.
Due to the length of the current paper, we postpone such an investigation
to a subsequent study.

\section{Numerical implementation}

The curved boundary of the cylinder was approximated by polygons $\Omega_h$, where the edge lengths of $\partial\Omega_h$ are of order $h$ in size. 
Then conventional finite elements can be employed, with the various boundary expressions being approximated by appropriate quantities. 
For the computations described in section \ref{sec:compapp}, we used the Robin-type
technique \cite{lrsBIBjm} together with the Scott--Vogelius elements of degree 4. The order of approximation for the numerical method is $h^{7/2}$
in the gradient norm.

The remaining results were computed using the lowest-order Taylor--Hood approximation.
To implement the Navier-slip boundary condition,
we used Nitsche's method \cite{lrsBIBiy,ref:StenbergNitscheLagrange,winter2018nitsche}
to enforce slip conditions in the limit of small mesh size.
The details regarding numerical implementation of \eqref{eqn:effnavstot}
together with boundary conditions \eqref{eqn:bceesnavst} and \eqref{eqn:navierslip},
are given in \cite{lrsBIBiy}.
The boundary integrals are approximated to order $h^2$, but the order of
approximation for the numerical method is only of order $h^{3/2}$
in the gradient norm.

\omitit{
\begin{remark}[Remark to editor and reviewers] \textit{
Before publication we will provide Jupyter notebooks for the students where they can interactively run the codes used for all of the examples. For the simulations in Section \ref{sec:compapp}, the numerical methods are implemented using FEniCS. In FEniCS, there is a close similarity between the variational formulations and the code, which makes it highly appropriate for educational purposes. However, the installation of FEniCS may pose a substantial bottleneck for students and faculty that are not familiar with Python. For this reason, we are looking into making a Jupyter hub where the code itself is run on a remote server. This would allow students to quickly get the code running.}
\end{remark}
}

\section{Conclusions}

We have shown that examining the Stokes paradox from different angles enriches
the understanding of the phenomenon.
The approaches dovetail together in the final analysis, but they allow
answers to different questions related to the paradox.
Perhaps the most critical question relates to what goes wrong when we pose
the Stokes problem on larger and larger domains.
We explored two different ways to consider this question, via numerical simulation
for general domains and analytical solutions on specific domains.
Fortunately, they give the same advice as to what happens in the limit,
and this agrees with the functional analysis formulation of the problem
on an infinite domain.
We showed that the Stokes paradox can arise in other flow problems as well.

\omitit{
\section{Future directions}

Fluid flow continues to be an active area of research.
A review of all that is going on today would take many pages,
so we mention only one area that is developing rapidly.

One thing that makes fluid flow interesting (and challenging)
is its instabilities \cite{helal2012benjamin}.
One paradox is that one of the simplest flows,
Couette flow \cite{ref:romanovstabilityCouette}, exhibits no linear instabilities
of the sort studied extensively \cite{chandrasekhar2013hydrodynamic}.
But this paradox has recently been resolved \cite{lrsBIBis}
by considering a more general definition of instability.
This type of instability has also been studied recently in other flows \cite{lrsBIBiw}.
}

\section{Acknowledgments}

We thank Vivette Girault for valuable information and advice.
\bibliographystyle{plain}

\end{document}